\documentclass[12pt]{article}

\usepackage{scicite}
\usepackage{times}
\usepackage{cite}
\usepackage{url}
\usepackage{amssymb,amsmath,amsthm}
\usepackage{bm}
\usepackage{graphicx}
\usepackage{enumerate}
\usepackage{microtype}
\usepackage{color}
\usepackage{hyperref}
\usepackage{multirow}
\usepackage{multirow}

\theoremstyle{plain}
\newtheorem*{theorem*}{Theorem}
\newtheorem{theorem}{Theorem}

\definecolor{webgreen}{rgb}{0,.35,0}
\definecolor{webbrown}{rgb}{.6,0,0}
\definecolor{RoyalBlue}{rgb}{0,0,0.9}
\definecolor{purp}{rgb}{0.6,0.05,0.8}
\definecolor{ora}{rgb}{0.7,0.35,0.02}

\hypersetup{
   colorlinks=true, linktocpage=true, 
   urlcolor=webbrown, linkcolor=RoyalBlue, citecolor=webgreen,
   pdfauthor={Lucy Liu, Gary P. T. Choi, L. Mahadevan},
   pdfsubject={Wallpaper group kirigami}
}

\newcounter{lastnote}

\begin{document}
\author{Lucy Liu$^{1\dagger}$, Gary P. T. Choi$^{2\dagger}$, L. Mahadevan$^{3,4\ast}$\\
\\
\footnotesize{$^{1}$Harvard College, Cambridge, MA, USA}\\
\footnotesize{$^{2}$Department of Mathematics, Massachusetts Institute of Technology, Cambridge, MA, USA}\\
\footnotesize{$^{3}$School of Engineering and Applied Sciences, Harvard University, Cambridge, MA, USA}\\
\footnotesize{$^{4}$Departments of Physics, and Organismic and Evolutionary Biology, Harvard University, Cambridge, MA, USA}\\
\footnotesize{$^{\dagger}$These authors contributed equally to this work.}\\
\footnotesize{$^\ast$To whom correspondence should be addressed; E-mail: lmahadev@g.harvard.edu}
}
\title{Wallpaper group kirigami}
\date{} 

\baselineskip24pt

\maketitle

\begin{abstract}
Kirigami, the art of paper cutting, has become a paradigm for mechanical metamaterials in recent years. The basic building blocks of any kirigami structures are repetitive deployable patterns that derive inspiration from geometric art forms and simple planar tilings. Here we complement these approaches by directly linking kirigami patterns to the symmetry associated with the set of seventeen repeating patterns that fully characterize the space of periodic tilings of the plane. We start by showing how to construct deployable kirigami patterns using any of the wallpaper groups, and then design symmetry-preserving cut patterns to achieve arbitrary size changes via deployment. We further prove that different symmetry changes can be achieved by controlling the shape and connectivity of the tiles and connect these results to the underlying kirigami-based lattice structures. All together, our work provides a systematic approach for creating a broad range of kirigami-based deployable structures with any prescribed size and symmetry properties.
\end{abstract}

\section{Introduction}
Kirigami, the creative art of paper cutting, has recently transformed from a beautiful art form into a promising approach for the science and engineering of shape and thence function. By introducing architected cuts into a thin sheet of material, one can achieve deployable structures with auxetic properties while morphing into pre-specified shapes. This has led to a number of studies on the geometry, topology and mechanics of kirigami structures~\cite{chen2016topological,tang2017design,lubbers2019excess,blees2015graphene,shyu2015kirigami}. Most of these studies start with a relatively simple set of basic building blocks of kirigami patterns that take the form of triangles~\cite{grima2006auxetic} or quads~\cite{grima2000auxetic}, although on occasion they take inspiration from art in the form of ancient Islamic tiling patterns~\cite{rafsanjani2016bistable}, which are periodic. The periodicity of the pattern allows us to easily scale up the design of a deployable structure without changing its overall shape. Recently, there have been attempts to explore generalizations of the cut geometry~\cite{choi2019programming,choi2020compact} and cut topology~\cite{chen2020deterministic} moving away from purely periodic deployable kirigami base patterns. However, it is still unclear how one might explore such base patterns systematically. Since the deployment of a kirigami structure is largely driven by the local rotation of the tiles, it is natural to ask what class of symmetries and size changes of the deployed structure can be achieved by controlling the tile geometry and connectivity.

A natural place to begin in our quest to address this question is to turn to the class of two-dimensional repetitive patterns that tile the plane, which are characterized by the plane crystallographic groups (the \emph{wallpaper groups})~\cite{grunbaum1986tilings}. A remarkable result by Fedorov~\cite{fedorov1891simmetrija} and P{\'o}lya~\cite{polya1924analogie} is that there are exactly seventeen distinct wallpaper groups with different properties in terms of the rotational, reflectional, and glide reflectional (i.e. the combination of a reflection over a line and a translation along the line) symmetries. Furthermore, the crystallographic restriction theorem tells us that the order of rotational symmetry in any wallpaper group pattern can only be $n = 1, 2, 3, 4, 6$. Table~\ref{table:wallpaper_group} lists the seventeen wallpaper groups (represented using the crystallographic notations) with their symmetry properties~\cite{radaelli2011symmetry}. While wallpaper groups have started to form the basis for planar electromagnetic metamaterials~\cite{padilla2007group,bingham2008planar} and topology optimization~\cite{mao2020designing}, they do not seem to have been explored in the context of kirigami-based mechanical metamaterials, with only a few patterns identified~\cite{stavric2019geometrical}. Here, we remedy this and consider all 17 of the wallpaper groups for the design of deployable kirigami patterns.

\begin{table}[t]
    \centering
    \begin{tabular}{c|cccccc|cc}
     \textbf{Rotational} & \multicolumn{8}{c}{\textbf{Reflectional symmetry}} \\ \cline{2-9}
     \textbf{symmetry} & \multicolumn{6}{c|}{\textbf{Yes}} & \multicolumn{2}{c}{\textbf{No}}\\ \hline
    6-fold & \multicolumn{6}{c|}{p6m} & \multicolumn{2}{c}{p6}\\ \hline
    \multirow{2}{*}{4-fold} & \multicolumn{6}{c|}{\textbf{Any mirrors at $45^{\circ}$?}} & \multicolumn{2}{c}{\multirow{2}{*}{p4}}\\ \cline{2-7}
    & \multicolumn{3}{c|}{Yes: p4m} & \multicolumn{3}{c|}{No: p4g} & \\ \hline
    \multirow{2}{*}{3-fold} & \multicolumn{6}{c|}{\textbf{Any rotation center off mirrors?}} & \multicolumn{2}{c}{\multirow{2}{*}{p3}}\\ \cline{2-7}
    & \multicolumn{3}{c|}{Yes: p31m} & \multicolumn{3}{c|}{No: p3m1} & \\ \hline
    \multirow{3}{*}{2-fold} & \multicolumn{6}{c|}{\textbf{Any perpendicular reflections?}} & \multicolumn{2}{c}{\textbf{Any glide reflection?}}\\ \cline{2-9}
    & \multicolumn{4}{c|}{\textbf{Yes; Any rotation center off mirrors?}} & \multicolumn{2}{c|}{No:} & \multicolumn{1}{c|}{\multirow{2}{*}{Yes: pgg}} & \multirow{2}{*}{No: p2} \\ \cline{2-5}
    & \multicolumn{2}{c|}{Yes: cmm} & \multicolumn{2}{c|}{No: pmm} & \multicolumn{2}{c|}{pmg} & \multicolumn{1}{c|}{ } & \\ \hline
    1-fold & \multicolumn{6}{c|}{\textbf{Any glide reflection axis off mirrors?}} & \multicolumn{2}{c}{\textbf{Any glide reflection?}}\\ \cline{2-9}
    (none) & \multicolumn{3}{c|}{Yes: cm} & \multicolumn{3}{c|}{No: pm} & \multicolumn{1}{c|}{Yes: pg} & No: p1  \\ 
    \end{tabular}
    \caption{Characterization of the seventeen wallpaper groups~\cite{radaelli2011symmetry}.}
    \label{table:wallpaper_group}
\end{table}

\begin{figure}[!t]
\centering
\includegraphics[width=\textwidth]{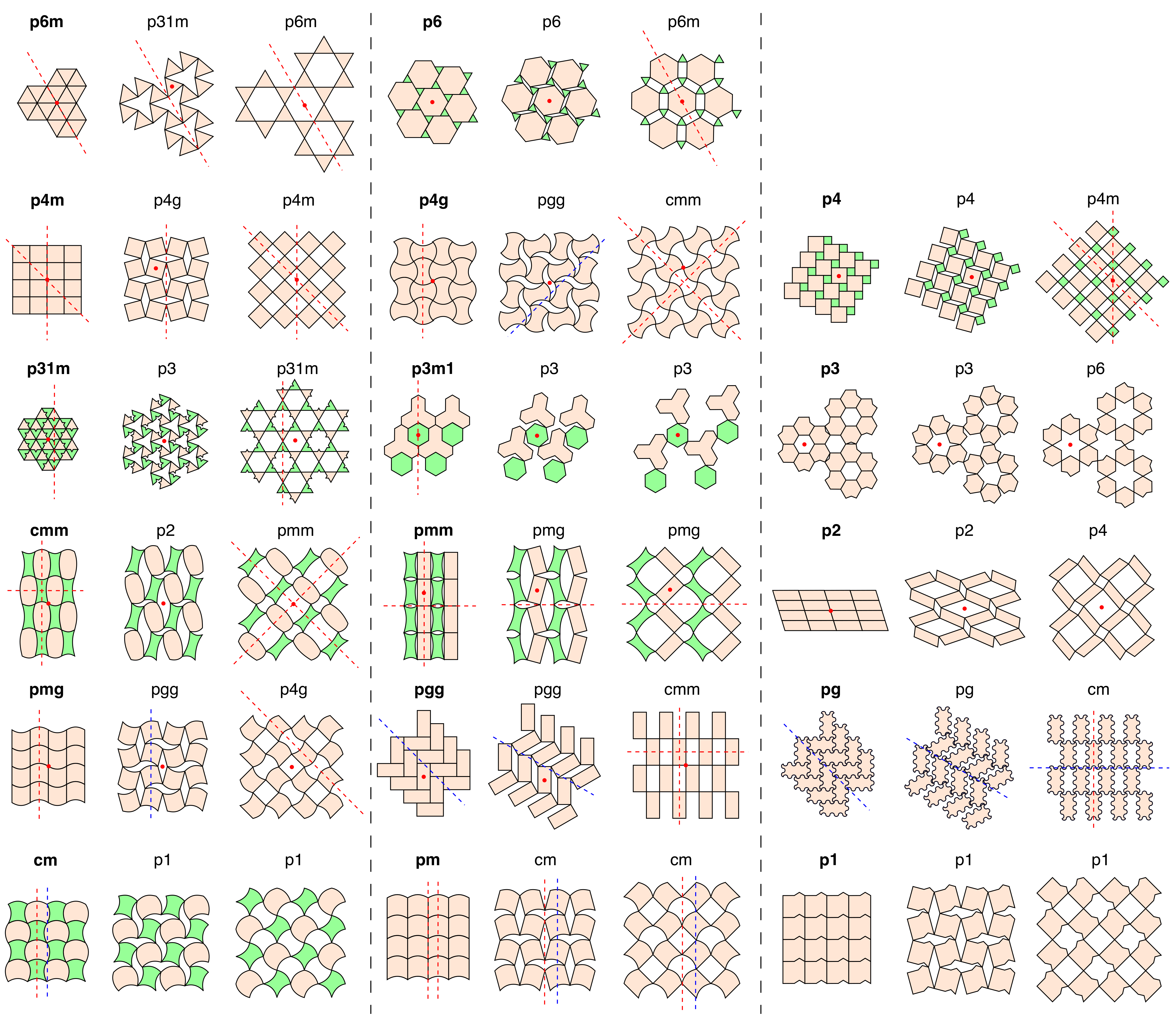}
\caption{\textbf{Examples of periodic deployable kirigami patterns in the seventeen wallpaper groups}. For each example (with the crystallographic notation boldfaced), we show a portion of the initial contracted state, an intermediate deployed state and the fully deployed state. Tiles with different shapes are in different colors. Key reflection axes (red dotted lines), glide reflection axes (blue dotted lines) and rotation centers (red dots) that can be used for determining their wallpaper group type are highlighted.}
\label{fig:F1}
\end{figure}

\section{Existence of deployable wallpaper group patterns}
The first question that naturally arises is whether all seventeen wallpaper groups can be used for designing deployable kirigami patterns. We answer this question by establishing the following result:
\begin{theorem}\label{thm:existence}
For any group $G$ among the seventeen wallpaper groups, there exists a deployable kirigami pattern in~$G$.
\end{theorem}
\noindent \textbf{Proof. } We prove this result by constructing explicit examples of periodic deployable structures in all seventeen wallpaper groups (Fig.~\ref{fig:F1}). Key reflection axes, glide reflection axes and rotation centers are highlighted and can be used together with Table~\ref{table:wallpaper_group} for determining the wallpaper group type for each of them. The result follows immediately from the existence of these patterns.\hfill $\blacksquare$\\

\begin{figure}[t!]
    \centering
    \includegraphics[width=\textwidth]{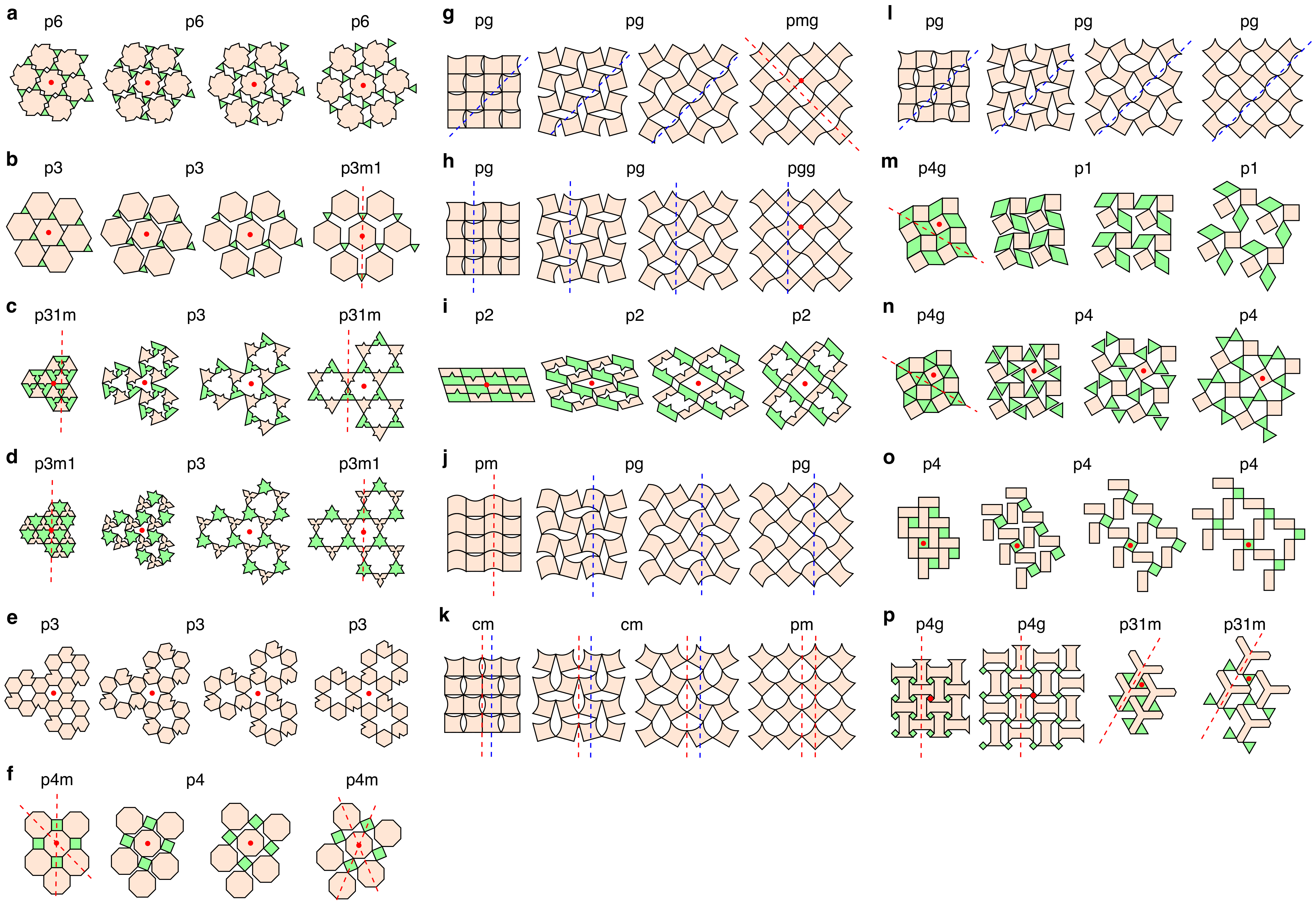}
    \caption{\textbf{More examples of periodic deployable patterns.} \textbf{a}-\textbf{b}, Two rigid-deployable patterns derived from the p6 example in Fig.~\ref{fig:F1}. \textbf{c}-\textbf{e}, Three rigid-deployable patterns derived from the standard kagome pattern. \textbf{f}-\textbf{l}, Seven rigid-deployable patterns derived from the standard quad pattern. \textbf{m}, A rigid-deployable p4g pattern consisting of squares and rhombi. Note that the pattern becomes p1 once deployed. \textbf{n}, Another rigid-deployable p4g pattern created by breaking the rhombi in \textbf{m} into triangles. This time, the pattern becomes p4 throughout the deployment. \textbf{o}, A rigid-deployable p4 pattern. Note that it has the same underlying topology as \textbf{n}. \textbf{p},~Two bistable Islamic tiling patterns~\cite{rafsanjani2016bistable} which are not rigid-deployable. Geometrical frustration exists at the intermediate deployments, while the initial and final states shown are frustration-free. Key examples of the reflection axes (red dotted lines), glide reflection axes (blue dotted lines) and rotation centers (red dots) that can be used for determining their wallpaper group type are highlighted.}
    \label{fig:SI_patterns}
\end{figure}

Note that all patterns in Fig.~\ref{fig:F1} are rigid-deployable, i.e. there is no geometrical frustration in the deployment of them (see also SI Video~S1). More examples of periodic deployable patterns are given in Fig.~\ref{fig:SI_patterns}. Fig.~\ref{fig:SI_patterns}\textbf{a}-\textbf{b} show two rigid-deployable patterns derived from the p6 example in Fig.~\ref{fig:F1}. Fig.~\ref{fig:SI_patterns}\textbf{c}-\textbf{e} show three rigid-deployable patterns derived from the standard kagome pattern. Fig.~\ref{fig:SI_patterns}\textbf{f}-\textbf{l} show seven patterns derived from the standard quad pattern. Fig.~\ref{fig:SI_patterns}\textbf{m}-\textbf{n} show two rigid-deployable p4g patterns, with different underlying topologies that lead to different wallpaper group changes under deployment. Fig.~\ref{fig:SI_patterns}\textbf{o} shows a rigid-deployable p4 pattern with the same topology as the pattern in Fig.~\ref{fig:SI_patterns}\textbf{n}. It is noteworthy that not all deployable kirigami patterns are rigid-deployable. Fig.~\ref{fig:SI_patterns}\textbf{p} shows two bistable p4g and p3m1 Islamic tiling patterns~\cite{rafsanjani2016bistable}, which exhibit geometrical frustration at the intermediate states of the deployment while being frustration-free at the contracted and final deployed states. Note that Theorem~\ref{thm:existence} focuses on the initial (contracted) state of deployable kirigami patterns. In fact, from Fig.~\ref{fig:F1} and Fig.~\ref{fig:SI_patterns}, we also have the following result:

\begin{theorem}\label{thm:existence_deployed}
For any wallpaper group $G$ among the seventeen wallpaper groups, there exists a deployable kirigami pattern with its final deployed shape in~$G$.
\end{theorem}
\noindent \textbf{Proof. } We prove the result by explicitly constructing examples of periodic deployable patterns with final deployed shape in any of the seventeen wallpaper groups:
\begin{itemize}
    \item p6m: See the p6m $\to$ p31m $\to$ p6m example in Fig.~\ref{fig:F1}.
    \item p6: See the p6 $\to$ p6 $\to$ p6 example in Fig.~\ref{fig:SI_patterns}\textbf{a}.
    \item p4m: See the p4m $\to$ p4g $\to$ p4m example in Fig.~\ref{fig:F1}.
    \item p4g: See the pmg $\to$ pgg $\to$ p4g example in Fig.~\ref{fig:F1}.
    \item p4: See the p2 $\to$ p2 $\to$ p4 example in Fig.~\ref{fig:F1}.
    \item p31m: See the p31m $\to$ p3 $\to$ p31m example in Fig.~\ref{fig:SI_patterns}\textbf{c}.
    \item p3m1: See the p3m1 $\to$ p3 $\to$ p3m1 example in Fig.~\ref{fig:SI_patterns}\textbf{d}.
    \item p3: See the p3m1 $\to$ p3 $\to$ p3 example in Fig.~\ref{fig:F1}.
    \item cmm: See the pgg $\to$ pgg $\to$ cmm example in Fig.~\ref{fig:F1}.
    \item pmm: See the cmm $\to$ p2 $\to$ pmm example in Fig.~\ref{fig:F1}.
    \item pmg: See the pg $\to$ pg $\to$ pmg example in Fig.~\ref{fig:SI_patterns}\textbf{g}.
    \item pgg: See the pg $\to$ pg $\to$ pgg example in Fig.~\ref{fig:SI_patterns}\textbf{h}.
    \item p2: See the p2 $\to$ p2 $\to$ p2 example in Fig.~\ref{fig:SI_patterns}\textbf{i}.
    \item cm: See the pg $\to$ pg $\to$ cm example in Fig.~\ref{fig:F1}.
    \item pm: See the cm $\to$ cm $\to$ pm example in Fig.~\ref{fig:SI_patterns}\textbf{k}.
    \item pg: See the pg $\to$ pg $\to$ pg example in Fig.~\ref{fig:SI_patterns}\textbf{l}.
    \item p1: See the cm $\to$ p1 $\to$ p1 example in Fig.~\ref{fig:F1}.
\end{itemize}
\hfill $\blacksquare$\\

\section{Size change throughout deployment}
\begin{figure}[!t]
\centering
\includegraphics[width=\textwidth]{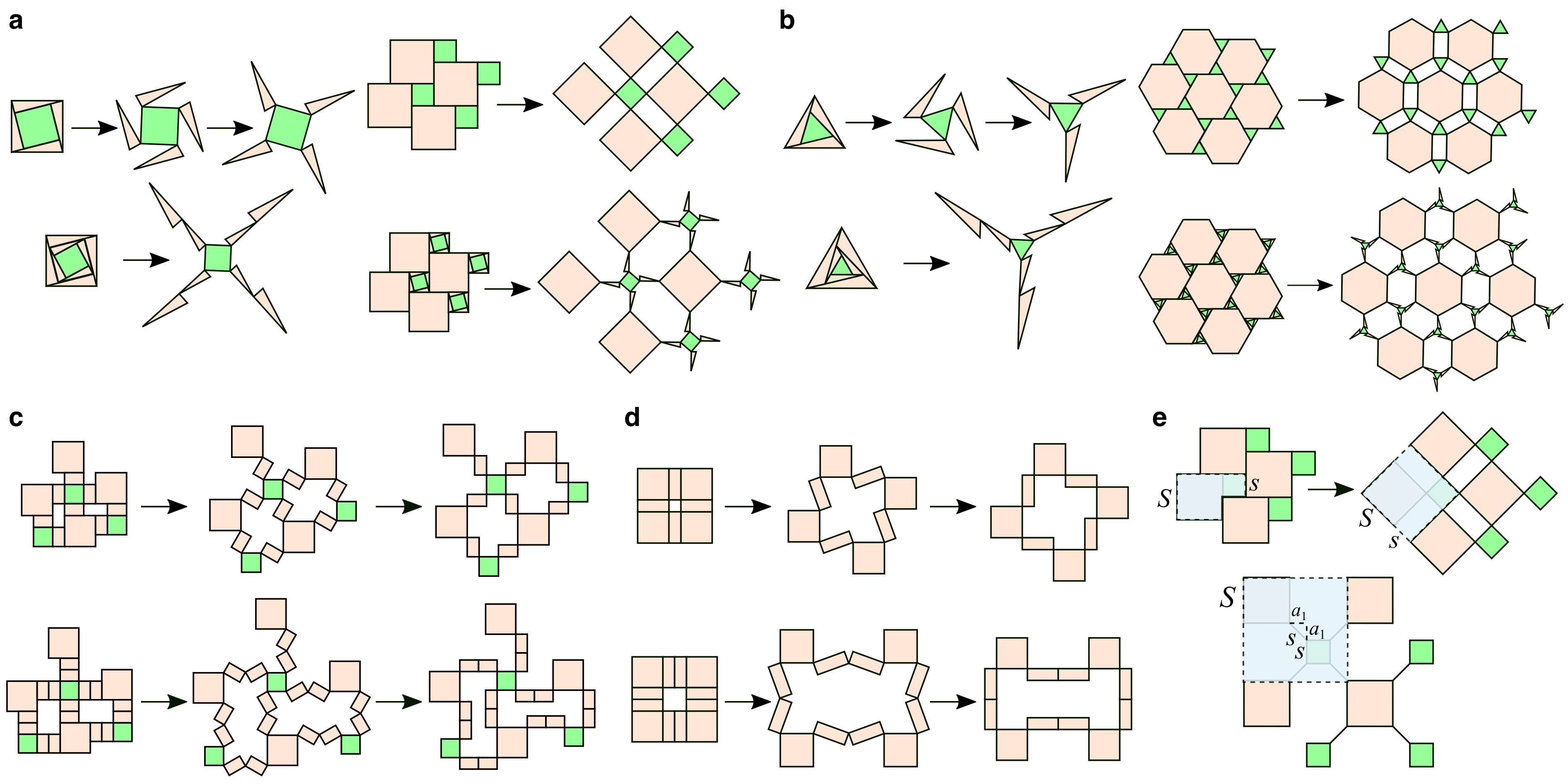}
\caption{\textbf{Symmetry-preserving expansion}. \textbf{a}, An expansion cut pattern on a square with 4-fold rotational symmetry (top left). The pattern can be refined hierarchically to achieve a larger size change (bottom left). These expansion cut patterns can be utilized for augmenting deployable patterns with $1$-,$2$-, or $4$-fold rotational symmetry, such as the p4 pattern in Fig.~\ref{fig:F1}, to achieve an arbitrary size change while preserving the rotational symmetry (right). \textbf{b}, An expansion cut pattern on a regular triangle with 3-fold rotational symmetry (top left). The pattern can be refined hierarchically to achieve a larger size change (bottom left). These expansion cut patterns can be utilized for augmenting deployable patterns with $1$-, $3$-, or $6$-fold rotational symmetry, such as the p6 pattern in Fig.~\ref{fig:F1} to achieve an arbitrary size change while preserving the rotational symmetry (right). \textbf{c},~Another type of expansion cuts on a p4 pattern produced by placing additional rectangular units between tiles. The first row shows the contracted, intermediate and fully deployed state of an augmented p4 pattern with 1 expansion layer. The second row shows the contracted and deployed state of an augmented p4 pattern with 2 expansion layers. \textbf{d},~An augmented p4m pattern constructed in a similar manner. \textbf{e},~The top row shows the contracted and deployed state of a deployable p4 pattern, with the shaded blue regions representing a unit cell and its deployed shape. The bottom row shows an augmented version of it with 1 level of ``ideal'' expansion cuts of infinitesimal width.}
\label{fig:F2}
\end{figure}

After showing the existence of deployable kirigami patterns in all seventeen wallpaper groups for both the contracted and deployed states, it is natural to ask whether some of the wallpaper groups are more advantageous over the others in terms of deployable kirigami design. In particular, one may wonder whether the size change of a deployable pattern is limited by its symmetry. It is clear that the size change is not limited by the reflectional symmetry or the glide reflectional symmetry. Here we show that the size change can in fact be arbitrary for any given rotational symmetry:
\begin{theorem}\label{thm:size}
For any deployable wallpaper group pattern with $n$-fold rotational symmetry, we can design an associated pattern with $n$-fold rotational symmetry and arbitrary size change.
\end{theorem}
The result is achieved by designing certain expansion methods for augmenting a given pattern with $n$-fold symmetry without breaking its symmetry. Two expansion methods are introduced below (see Fig.~\ref{fig:F2} and SI Video S2).

\subsection{Symmetry-preserving expansion cuts}
To achieve significant size change while preserving rotational symmetry, expansion cuts can be introduced to select rotating units in the pattern. Using a 4-fold expansion cut on a square in a 1-fold, 2-fold or 4-fold pattern, we can achieve an expansion of the pattern without changing its rotational symmetry (Fig.~\ref{fig:F2}\textbf{a}). Using a 3-fold expansion cut on a triangle in a pattern with 1-fold, 3-fold or 6-fold rotational symmetry, we can achieve an expansion of the pattern without changing its rotational symmetry (Fig.~\ref{fig:F2}\textbf{b}). 

\begin{figure}[!t]
    \centering
    \includegraphics[width=\textwidth]{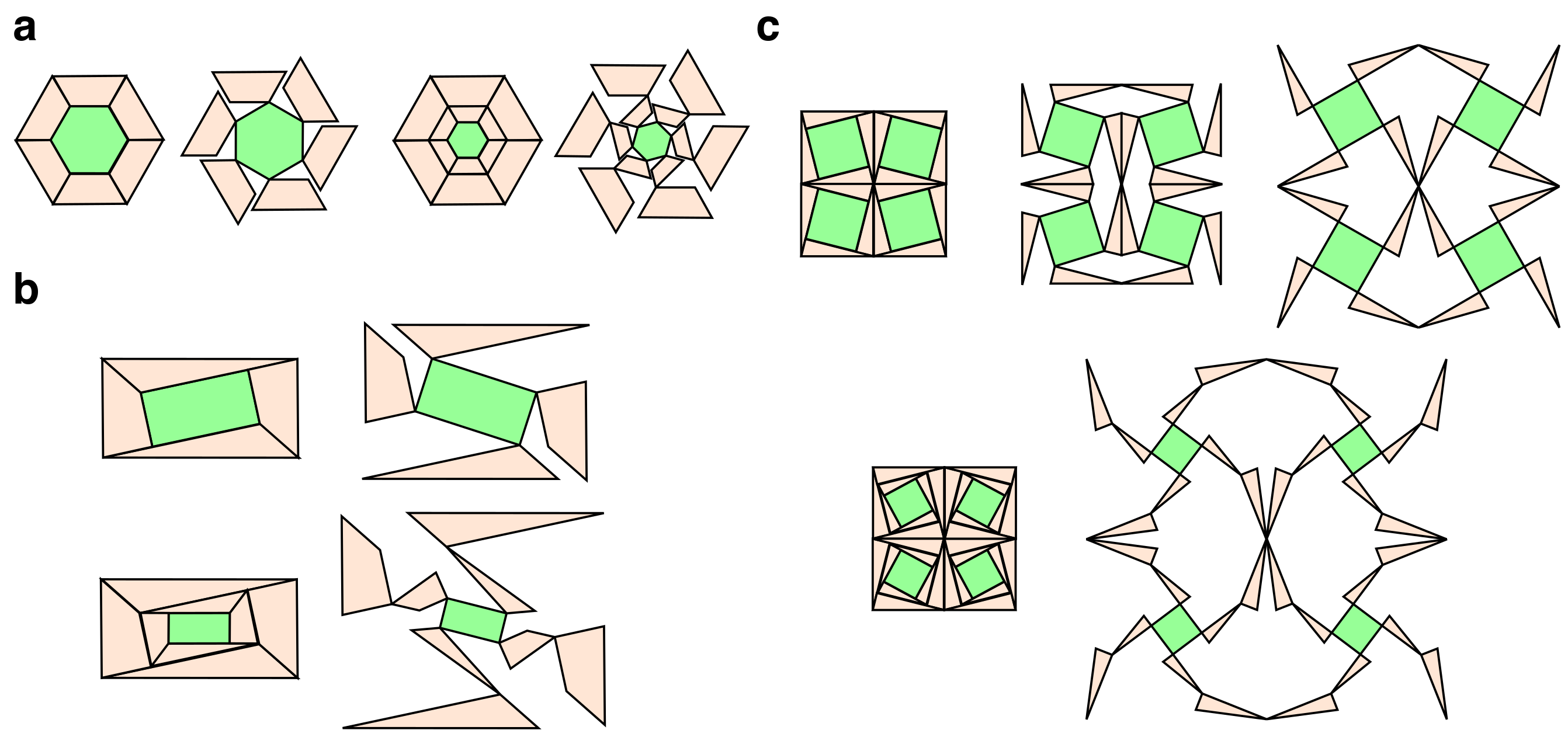}
    \caption{\textbf{More symmetry-preserving expansion cut patterns.} \textbf{a}, An expansion cut pattern that can be introduced on any tile with 6-fold rotational symmetry. The cut pattern achieves an expansion throughout deployment while preserving the 6-fold symmetry of the tile. \textbf{b}, An expansion cut pattern that can be introduced on any tile with 2-fold rotational symmetry. The cut pattern achieves an expansion throughout deployment while preserving the 2-fold symmetry of the tile. \textbf{c}, An expansion cut pattern with 2-fold rotational symmetry that preserves both the rotational symmetry and reflectional symmetry. The expansion cut pattern is derived from the pattern in Fig.~\ref{fig:F2}\textbf{a}, with four copies of it placed appropriately to form a pattern that preserves the reflectional symmetry throughout the deployment. The top row shows the deployment of the pattern with 1 level of cuts. The bottom row shows the deployment of the pattern with 2 levels of cuts.}
    \label{fig:SI_expansion_reflection}
\end{figure}

While the above expansion cuts are introduced on a square and a triangle only, it is easy to see that similar expansion cuts can be introduced on any tiles with 4-fold and 3-fold rotational symmetry respectively. Fig.~\ref{fig:SI_expansion_reflection}\textbf{a} shows a symmetry-preserving expansion cut pattern that can be introduced on any 6-fold tile (e.g. a regular hexagon). The cut pattern preserves the 1-fold, 2-fold, 3-fold or 6-fold rotational symmetry of the entire kirigami pattern. Fig.~\ref{fig:SI_expansion_reflection}\textbf{b} shows a symmetry-preserving expansion cut pattern that can be introduced on any 2-fold tile (e.g. a rectangle). The cut pattern preserves the 1-fold or 2-fold rotational symmetry of the entire kirigami pattern. It can be observed that by increasing the level of cuts, we can achieve a larger size change.

Note that the pattern in Fig.~\ref{fig:F2}\textbf{a} can be utilized for augmenting a given periodic deployable pattern to achieve an arbitrary size change, while the reflectional symmetry of the given pattern may be lost. Fig.~\ref{fig:SI_expansion_reflection}\textbf{c} shows an expansion cut pattern with 2-fold rotational symmetry derived from it. We suitably reflect the pattern to form an expansion cut pattern on a square consisting of 16 triangles and 4 squares. Note that the new cut pattern is not only with 2-fold rotational symmetry but also reflectional symmetry. Therefore, it can be utilized for augmenting a given periodic deployable pattern with $1$- or $2$-fold rotational symmetry while preserving both its rotational symmetry and reflectional symmetry.

\subsection{Symmetry-preserving expansion tiles}
Another way to design an associated pattern with increased size change is to add rotating units between adjacent tiles of the original pattern (Fig.~\ref{fig:F2}\textbf{c}-\textbf{d}). More specifically, we augment a given deployable pattern by adding thin rectangles between adjacent tiles, which allow for greater expansion when the pattern is deployed. Analogous to the above-mentioned method, it is possible to preserve the rotational symmetry of the given pattern by appropriately placing the additional units. Again, it is possible to preserve the reflectional symmetry of the contracted state or even the deployed state of certain patterns using this method (for example, the pattern in Fig.~\ref{fig:F2}\textbf{d} with an even number of expansion layers). We remark that this method introduces gaps to the contracted state of the new pattern. 

\subsection{Analysis of the size change}
To quantify the size change achieved by our proposed symmetry-preserving expansions, we consider the p4 pattern in Fig.~\ref{fig:F2}\textbf{e} and denote the side length of the larger and smaller squares in the original pattern as $S$ and $s$, with $s \leq S$. We measure the size change of the pattern upon deployment by selecting a unit cell in the contracted state and comparing its area to that of a corresponding unit cell in the deployed state (Fig.~\ref{fig:F2}\textbf{e}, the shaded regions in the top row). It is easy to see that the contracted unit cell has area $S^2 + s^2$ and the deployed unit cell has area $(S + s)^2$. Therefore, the base size change ratio is 
\begin{equation}
    r_0 = \frac{(S+s)^2}{S^2 + s^2}.
\end{equation}
This ratio simplifies to $2$ when $S = s$ and $\frac{9}{5}$ when $S = 2s$. 

Each expansion cut creates a new unit that, upon deployment, rotates to further separate the original tiles of the base pattern. In the fully deployed state, we let $a_i$ be the additional vertical and horizontal separation introduced by each cut in the $i$-th round of expansion cuts. The expansion cuts also shave area off of the original tiles in order to form the new rotating units. Let $b_i$ be the width that the squares of side length $s$ lose from each cut in the $i$-th round of expansion cuts. 

With $n$ rounds of expansion cuts, the unit cell's area after deployment will be $(S + s + 2\sum_{i = 1}^n(a_i - b_i))^2 $ and the size change ratio will be 
\begin{equation}
    r_n = \frac{(S + s + 2\sum_{i = 1}^n(a_i - b_i))^2 }{S^2 + s^2}.
\end{equation}

The values of elements in $a_i$ and $b_i$ depend on the shape and width of expansion cuts. If we consider ``ideal'' expansion cuts of length $s$ and infinitesimal width, then $a_i = \frac{s}{\sqrt2}$ and $b_i = 0$ for all $i$ (Fig.~\ref{fig:F2}\textbf{e}, bottom row). For these ideal cuts, the size change ratio after $n$ rounds of expansion would be 
\begin{equation}
r_n = \frac{(S + s + 2n\frac{s}{\sqrt2})^2}{S^2 + s^2} = \frac{(S + s + \sqrt2 ns)^2}{(S^2 + s^2)}.
\end{equation}
This suggests that the size change ratio scales approximately with $n^2$, and we can achieve an arbitrary size change by choosing a sufficiently large $n$. 

\begin{figure}[t!]
    \centering
    \includegraphics[width=\textwidth]{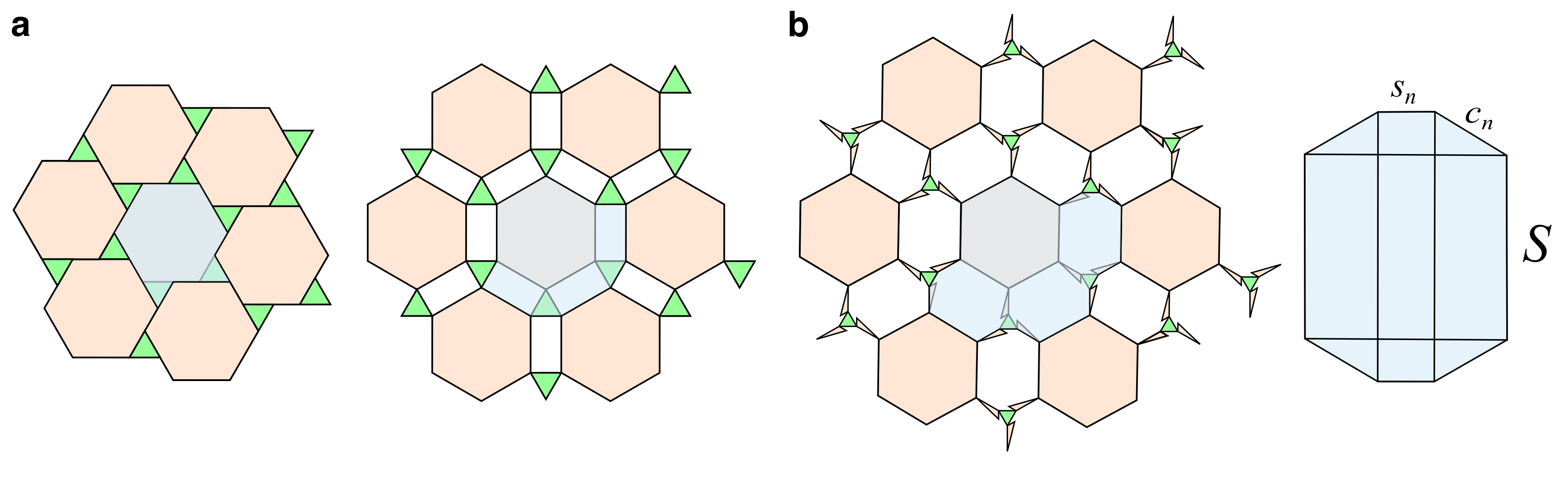}
    \caption{\textbf{Size change in the triangle-hexagon deployment pattern.} \textbf{a}, The contracted and deployed states of the pattern with no expansion cuts used. The shaded areas represent the unit cell used to calculate the pattern's size change ratio. \textbf{b}, On the left, an image of the triangle-hexagon pattern with one round of expansion cuts deployed. On the right, a close-up of an irregular octagon formed by deployment of the expansion cuts. The octagon is broken into the smaller triangles and rectangles we use to determine its area.}
    \label{fig:SI_expansion_triangle}
\end{figure}

Similarly, one can perform an analysis on the size change of the triangle expansion cut pattern in Fig.~\ref{fig:F2}\textbf{b}. We select a unit cell in the contracted state and compare it to the corresponding units in the deployed and expanded states. Unit cells are represented as shaded areas in Fig.~\ref{fig:SI_expansion_triangle}. Let $S$ be the side length of the hexagons and $s$ be the side length of the triangles; a regular hexagon will have area $\frac{3\sqrt3}{2}S^2$ and a regular triangle area $\frac{\sqrt3}{4}s^2$. 

The contracted state unit cell consists of one hexagon and two triangles, which together have area $\frac{3\sqrt3}{2}S^2 + \frac{\sqrt{3}}{2}s^2$. The deployed state unit cell has three additional rectangles, each with area $Ss$. Then the deployed unit cell area is $\frac{3\sqrt3}{2}S^2 + \frac{\sqrt{3}}{2}s^2 + 3Ss$, and the base size change ratio is 
\begin{equation}
r_0 = \frac{3\sqrt3S^2 + \sqrt{3}s^2 + 6Ss}{3\sqrt3S^2 + \sqrt{3}s^2}.
\end{equation}

For this pattern, expansion cuts as done in Fig.~\ref{fig:F2}\textbf{b} introduce gaps with area equivalent to that of the irregular octagons shown in blue in Fig.~\ref{fig:SI_expansion_triangle}\textbf{b}. We assume expansion cuts are done in a manner that preserves the equilateral triangle shape of the green tiles.

Let $a_i$ be the additional separation the $i$-th round of expansion cuts adds between each pair of adjacent triangles and hexagons, so each pair's closest vertices are now $c_n = \sum_{i = 1}^n a_i$ apart. Let $b_i$ be the side length each equilateral triangle loses in the $i$th round of expansion cuts, so $s_n = s - \sum_{i = 1}^n b_i$ is the triangle's remaining side length after $n$ rounds of expansion cuts.

Each octagon can be broken into smaller rectangles and triangles as seen in the figure: four $30-60-90$ triangles of area $\frac{a_n^2\sqrt3}{8}$, two rectangles of area $\frac{c_ns_n}{2}$, two rectangles of area $\frac{Sc_n\sqrt3}{2}$, and a center rectangle of area $Ss_n$. The triangles are $30-60-90$ because maximal deployment occurs when the triangles and hexagons of the original pattern are as separated as possible. This occurs when each edge between a triangle vertex and a hexagon vertex bisects both vertex angles. The octagon's total area will then be $\frac{c_n^2\sqrt3}{2} + c_ns_n +  Sc_n\sqrt3 + Ss_n$.

After $n$ rounds of expansion cuts, the expanded unit cell consists of a regular hexagon with side length $S$, two equilateral triangles with side length $s_n$, and three irregular octagons as described above. This unit cell has area $\frac{3\sqrt3}{2}S^2 + \frac{\sqrt3}{2}s_n^2 + 3(\frac{c_n^2\sqrt3}{2} + c_ns_n +  Sc_n\sqrt3 + Ss_n).$ Then the size change ratio is
\begin{equation}
    r_n = \frac{3\sqrt3S^2 + \sqrt3s_n^2 + 6(\frac{c_n^2\sqrt3}{2} + c_ns_n +  Sc_n\sqrt3 + Ss_n)}{3\sqrt3S^2 + \sqrt{3}s^2},
\end{equation}
where $s_n = s - \sum_{i = 1}^n b_i$ and $c_n = \sum_{i=1}^n a_i$.

Now, if we consider ``ideal'' expansion cuts of length $s$ and infinitesimal width, we have $a_i = s$ and $b_i = 0$ for all $i$. It follows that $c_n = ns$ and $s_n = s$. Therefore, with these ideal expansion cuts we have
\begin{equation}
    r_n = \frac{3\sqrt3S^2 + \sqrt{3}s^2 + 6(\frac{n^2s^2\sqrt3}{2} + ns^2 + Sns\sqrt3 + Ss)}{3\sqrt3S^2 + \sqrt{3}s^2},
\end{equation}
which scales approximately with $n^2$ and is unbounded. This shows that we can achieve an arbitrary size change using the triangle expansion cut pattern with suitable refinements.

We are now ready to prove Theorem~\ref{thm:size}.\\

\noindent \textbf{Proof of Theorem~\ref{thm:size}. } As described above, we have explicitly constructed symmetry-preserving expansion cut patterns for the 6-fold, 4-fold, 3-fold and 2-fold cases; the construction of an expansion cut pattern for the 1-fold case is straightforward. Also, we have shown that an arbitrary size change can be achieved by increasing the number of cuts and making them arbitrarily thin. For any deployable kirigami pattern with $n$-fold rotational symmetry in the contracted state (where $n=1,2,3,4,6$), we have the following cases:\\
Case (i): There is a center of $n$-fold rotation at the center of a tile in the contracted state. In this case, we can simply introduce $n$-fold symmetry-preserving expansion cuts in this tile as part of a unit cell in the repetitive pattern. \\
Case (ii): There is a center of $n$-fold rotation at either a vertex, the center of an edge, the center of a rift (i.e. a gap that forms when tiles separate during deployment), or the center of a void (i.e. a gap in between some tiles) in the contracted state. This implies that there are $n$ identical tiles around this rotation center in a unit cell of the repetitive pattern. We can then add expansion cuts or expansion tiles for each of the $n$ tiles while keeping them identical with respect to rotation about the center. \\
Thus we can always obtain a deployable pattern with $n$-fold rotational symmetry and arbitrary size change. \hfill $\blacksquare$\\

We note that the above symmetry-preserving expansion methods also allow us to achieve an arbitrary perimeter change.

\section{Symmetry change throughout deployment}
Now, we study how the kirigami patterns change in terms of the wallpaper groups throughout the deployment. More specifically, what are the possible symmetry changes throughout the deployment? We have the following result: 
\begin{theorem} \label{thm:group_change}
Gain, loss, and preservation of symmetry are all possible throughout the deployment of a kirigami pattern.
\end{theorem}
\noindent \textbf{Proof. } We prove this result by noting that from Fig.~\ref{fig:F1} and Fig.~\ref{fig:F4} we can observe different types of symmetry change as a pattern expands from its contracted state to its deployed state:
\begin{itemize}
    \item Rotational symmetry gained: pmg $\to$ pgg $\to$ p4g (permanent).
    \item Rotational symmetry lost: p4g $\to$ pgg $\to$ cmm (permanent), p6m $\to$ p31m $\to$ p6m (temporal).
    \item Rotational symmetry preserved: p6 $\to$ p6 $\to$ p6m.
    \item Reflectional symmetry gained: pgg $\to$ pgg $\to$ cmm (permanent).
    \item Reflectional symmetry lost: p3m1 $\to$ p3 $\to$ p3 (permanent), p4g $\to$ pgg $\to$ cmm (temporal).
    \item Reflectional symmetry preserved: p4m $\to$ p4g $\to$ p4m.
    \item Glide reflectional symmetry gained: p1 $\to$ p1 $\to$ p4m (permanent).
    \item Glide reflectional symmetry lost: cm $\to$ p1 $\to$ p1 (permanent), cmm $\to$ p2 $\to$ pmm (temporal).
    \item Glide reflectional symmetry preserved: pg $\to$ pg $\to$ cm.
\end{itemize}
\hfill $\blacksquare$\\

We remark that although some patterns preserve rotational, reflectional or glide reflectional symmetry, the rotation centers and reflection axes do not necessarily remain fixed. For instance, the pattern cmm $\to$ p2 $\to$ pmm has rotation centers off mirrors at the initial state, while all rotation centers lie on mirrors at the final deployed state. For the pattern pm $\to$ cm $\to$ cm, the number of reflection axes decreases throughout deployment, while the number of glide reflection axes remains unchanged. 

\subsection{Symmetry change for a fixed cut topology}\label{sect:quad}
Note that several pattern examples in Fig.~\ref{fig:F1} are with the standard quad kirigami topology, where unit cells containing four tiles arranged as seen in the p4m example in Fig.~\ref{fig:F1} are connected in a larger grid. It can be observed that these patterns exhibit 1-fold, 2-fold, and 4-fold rotational symmetry, and some of them can even achieve a 2-fold to 4-fold rotational symmetry change (the p2 and pmg examples). Are other rotational symmetry gains possible for patterns with this topology, like changes from 1-fold to 2-fold or 4-fold rotational symmetry?

\begin{figure}[!t]
\centering
\includegraphics[width=\textwidth]{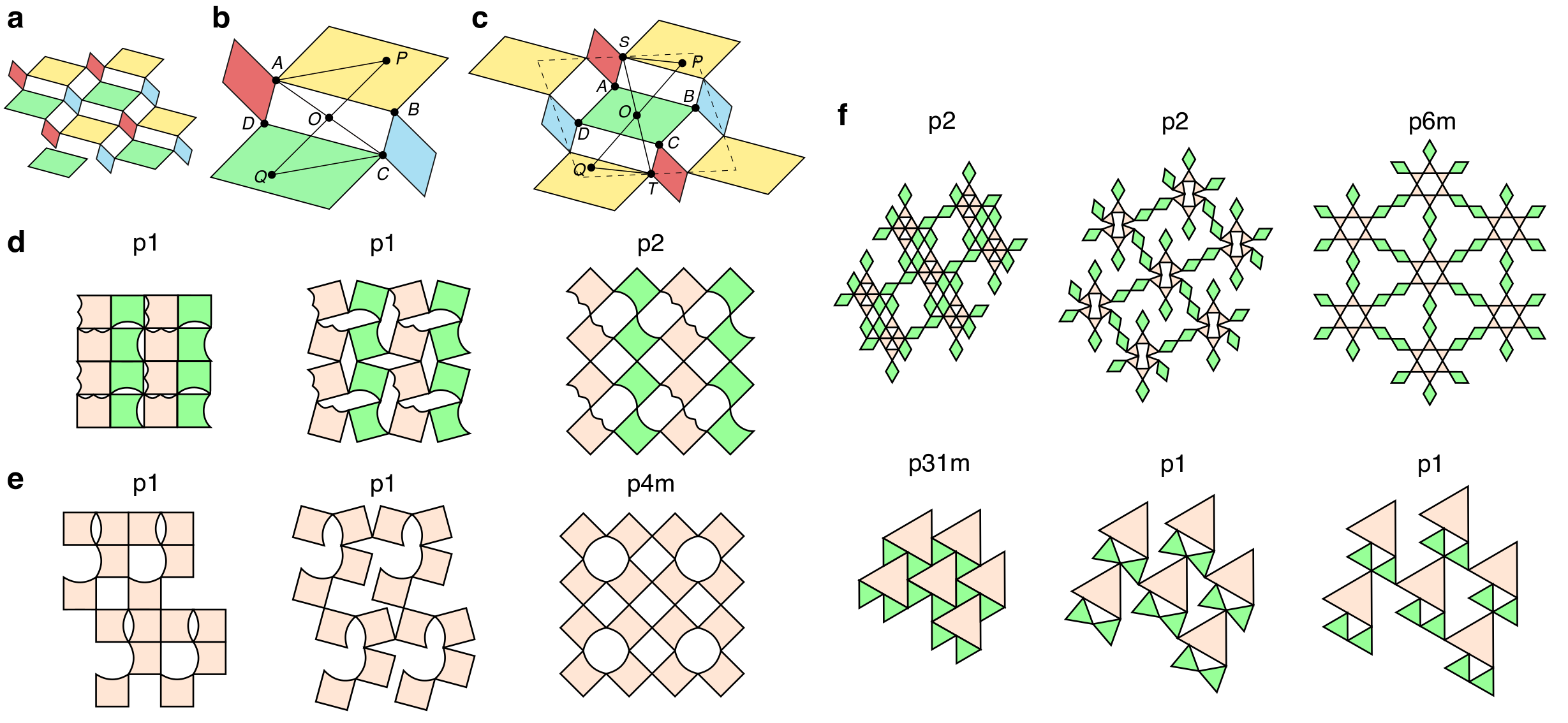}
\caption{\textbf{Exploring possible symmetry changes}. \textbf{a}, A general deployed quad pattern with 2-fold rotational symmetry. The parallelograms can be changed in pairs (red and blue, or yellow and green) to other shapes whose vertices form parallelograms. \textbf{b}, For the case where the center of rotation (COR) $O$ is at the center of a rift (i.e. a gap that forms when tiles separate during deployment) in the deployed state, consider two corresponding points $P$, $Q$ within tiles and denote the two opposite vertices of the rift by $A$ and $C$. We can show that $\Delta APO \cong \Delta CQO$ throughout deployment or contraction, which implies that the contracted pattern is also with 2-fold rotational symmetry. \textbf{c}, For the case where the COR $O$ is at the center of a tile. We remark that the pattern here should not be thought as a supercell of \textbf{b}; the parallelograms are just used for simplicity to represent the tiles and can be changed to other shapes. However, whether a COR is at the center of either a tile or a rift depends on the tile shapes. \textbf{d}, For the case where the COR $O$ is at the intersection point of two tiles, we construct a deployable pattern that achieves a 1-fold to 2-fold symmetry change. \textbf{e}, Using a variation of the standard quad kirigami topology, we can achieve a 1-fold to 4-fold symmetry change as well as a gain in reflection. \textbf{f},~Introducing floppiness can lead to a large variety of symmetry changes, such as a 2-fold to 6-fold symmetry gain (p2 $\to$ p2 $\to$ p6m) or a loss in all symmetries (p31m $\to$ p1 $\to$ p1). For each pattern, tiles with different shapes are in different colors.}
\label{fig:F4}
\end{figure}

Here we consider a general deployed pattern with the standard quad kirigami topology and 2-fold rotational symmetry (Fig.~\ref{fig:F4}\textbf{a}). Note that for simplicity we use parallelograms to represent the tiles. These parallelograms can be changed to other shapes so long as the four vertices where each shape connects to other tiles form parallelograms, and congruent parallelograms are formed by vertices of the red and blue shape pair, and the yellow and green shape pair. There are three possible cases for where $O$, a center of rotation (COR) of the deployed state, lies in a general deployed pattern: at the center of a rift, the center of a tile, or a point where two tiles connect. Below, we show that for the first two cases the pattern's contracted state must also have at least 2-fold rotational symmetry. For the third case, we find an example pattern whose contracted state has 1-fold rotational symmetry. 

\subsubsection*{(i) $O$ is at the center of a rift}
    In the unit cell containing the rift and its four adjacent tiles, consider any two points $P$ and $Q$ which lie within tiles and map to each other after a 180-degree rotation around $O$. Then $O$ is the midpoint of $PQ$, and we can construct the congruent triangles shown in Fig.~\ref{fig:F4}\textbf{b} involving $P, Q, O$, and two opposite vertices of the rift $A$ and $C$. $AO = OC$ because the center of a parallelogram bisects its diagonals. Since opposite parallelograms across the rift are congruent, the rift will be a parallelogram. 
    
    The rift changes shape during deployment, but the tiles themselves are rigid, so $P$ and $Q$ are a fixed translation from $A$ and $C$ respectively. Then $PA = CQ$ and $\angle PAB = \angle QCD$ at any point in deployment. Throughout deployment the rift remains a parallelogram, so $AO = OC$ and $\angle PAO = \angle QCO$. Therefore, $\Delta APO \cong \Delta CQO$ at all stages of deployment.
    
    It follows that $O$ remains a midpoint of $PQ$, so in the contracted state, a 180-degree rotation around $O$ still maps $P$ and $Q$ to each other. Then 2-fold rotational symmetry is preserved and $O$ remains a COR for the contracted unit cell. The full pattern is a grid of unit cells, and when we rotate it around $O$, each unit cell is rotated right onto another unit cell. As the unit cell has 2-fold rotational symmetry, the rotational symmetry of the entire pattern is preserved. Thus, the contracted pattern must have 2-fold rotational symmetry, so a 1-fold contracted state cannot deploy to a 2-fold or 4-fold state in this case.
    
\subsubsection*{(ii) $O$ is at the center of a tile}
    As shown in Fig.~\ref{fig:F4}\textbf{c}, we now consider the outlined unit cell centered at $O$. Within this cell, by choosing two points $P$ and $Q$ which map to each other through 180-degree rotation, constructing congruent triangles, and applying the argument from case (i), we can again see that as the angles change through deployment, $P$ and $Q$ remain a 180-degree rotation around $O$ apart. Therefore, the unit cell retains 2-fold rotational symmetry around $O$. Once again, $O$ also remains a COR for the full pattern due to its grid structure. This shows that a 1-fold contracted state cannot deploy to a 2-fold or 4-fold state in this case.
    
\subsubsection*{(iii) $O$ is at the intersection point of two tiles}
    As shown in Fig.~\ref{fig:F4}\textbf{d}, we construct an explicit example of a deployable pattern with a 1-fold contracted state and a 2-fold deployed state (p1 $\to$ p1 $\to$ p2). Note that for this case 4-fold symmetry is not possible in the deployed state as a 90-degree rotation maps a tile to a rift.

We conclude that for patterns with the standard quad kirigami topology, 1-fold to 4-fold rotational symmetry gain is not possible, and that 1-fold to 2-fold gain is possible only in case (iii). This type of analysis offers a systematic way to understand how deployment affects pattern symmetry. 

\subsection{Symmetry change by controlling the cut topology}
Note that the above analysis has only focused on the standard quad kirigami topology. If we consider other cut topologies, we can achieve a larger variety of symmetry changes. For instance, using a variation of the standard quad kirigami topology, one can achieve a pattern with a 1-fold to 4-fold symmetry change and a gain in reflectional symmetry (Fig.~\ref{fig:F4}\textbf{e}).

For the patterns we have considered so far, each tile has at least two vertices connected to vertices of neighboring tiles, and so the motions of all tiles are interrelated. However, one can also consider changing the underlying topology of certain kirigami patterns such that some of the tiles have only one vertex connected to another tile, thereby increasing the floppiness of the patterns. Fig.~\ref{fig:F4}\textbf{f} shows two patterns with floppy rhombus or triangle tiles. During and after deployment, these floppy tiles have only one vertex's position determined and are free to rotate around that fixed vertex. The p2 $\to$ p2 $\to$ p6m pattern exhibits a 2-fold to 6-fold symmetry change and a gain in reflectional symmetry, while the p31m $\to$ p1 $\to$ p1 pattern loses all symmetries throughout deployment. 

\subsection{Analysis of the possible symmetry changes}
Now, we present a more detailed analysis of the possible symmetry changes in terms of the gain, loss and preservation of reflectional, glide reflectional and rotational symmetries.
\subsubsection{Gain of symmetry}
\begin{figure}[t!]
    \centering
    \includegraphics[width=\textwidth]{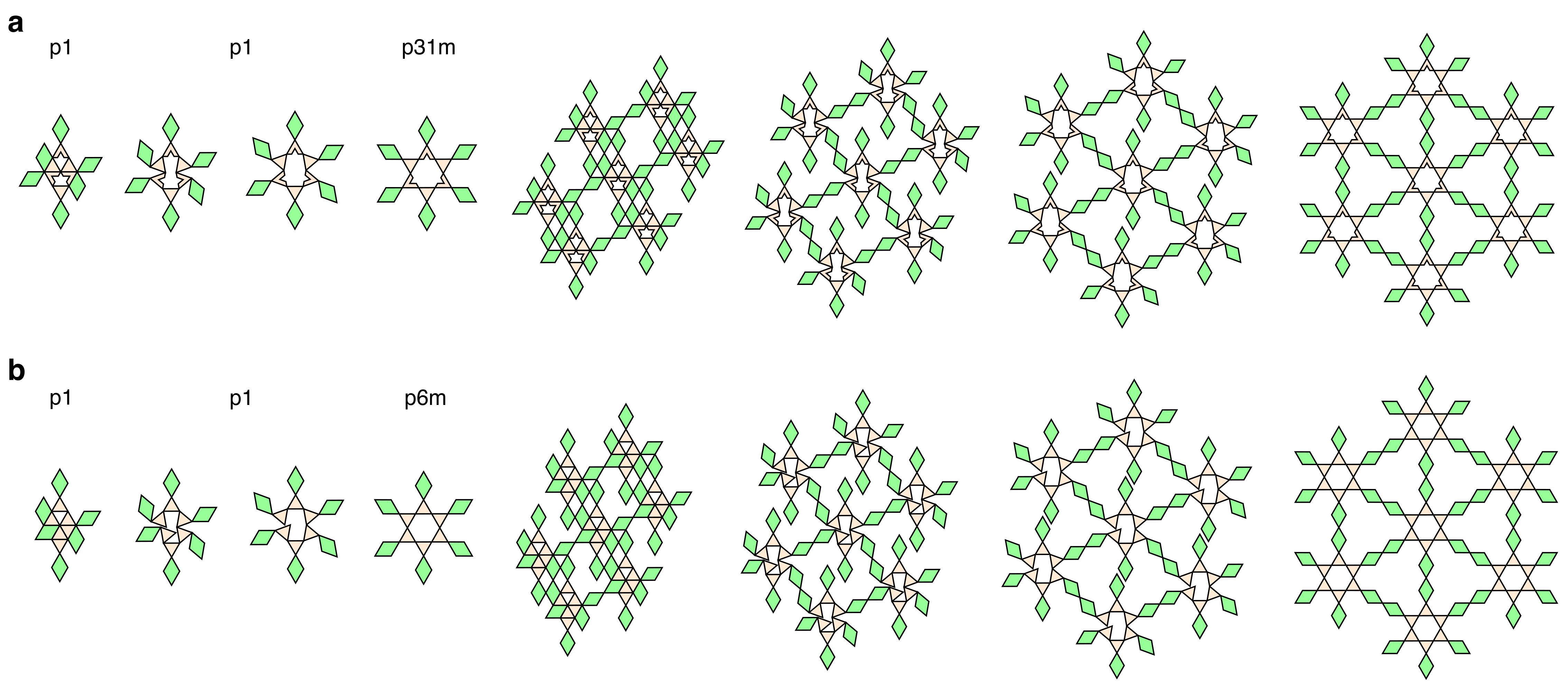}
    \caption{\textbf{More examples of deployable kirigami patterns with rotational symmetry gain throughout deployment.} \textbf{a}, A p1 $\to$ p1 $\to$ p31m example derived from the example in Fig.~\ref{fig:F4}\textbf{f}, with the shape of the triangles modified. Deployment of a unit cell is shown on the left, and deployment of a larger pattern section is shown on the right. \textbf{b}, A p1 $\to$ p1 $\to$ p6m example derived from the example in Fig.~\ref{fig:F4}\textbf{f}, with the geometry of the pattern and the connectivity of the triangles modified. Deployment of a unit cell is shown on the left, and deployment of a larger pattern section is shown on the right.}
    \label{fig:SI_symmetry_gain}
\end{figure}
\textbf{Rotational symmetry}: In Section~\ref{sect:quad}, we have explored the possible rotational symmetry gain for quad patterns. In fact, by considering more general periodic deployable patterns (possibly with floppy tiles), we can show that an $n$-fold to $m$-fold rotational symmetry gain is possible for any $n,m \in \{1,2,3,4,6\}$ with $n|m$ and $m > n$:
\begin{itemize}
    \item $1\to 2$: See the p1 $\to$ p1 $\to$ p2 example in Fig.~\ref{fig:F4}\textbf{d}, and the patterns in Fig.~\ref{fig:SI_patterns}\textbf{g}-\textbf{h}.
    \item $2\to 4$: See the pmg $\to$ pgg $\to$ p4g example and the p2 $\to$ p2 $\to$ p4 example in Fig.~\ref{fig:F1}.
    \item $3\to 6$: See the p3 $\to$ p3 $\to$ p6 example in Fig.~\ref{fig:F1}. 
    \item $1 \to 3$: See the p1 $\to$ p1 $\to$ p31m example in Fig.~\ref{fig:SI_symmetry_gain}\textbf{a}.
    \item $2 \to 6$: See the p2 $\to$ p2 $\to$ p6m example in Fig.~\ref{fig:F4}\textbf{f}. 
    \item $1 \to 4$: See the p1 $\to$ p1 $\to$ p4m example in Fig.~\ref{fig:F4}\textbf{e}.
    \item $1 \to 6$: See the p1 $\to$ p1 $\to$ p6m example in Fig.~\ref{fig:SI_symmetry_gain}\textbf{b}.
\end{itemize}

\textbf{Reflectional symmetry}: Gain of reflectional symmetry can be observed for all $n = 1,2,3,4,6$:
\begin{itemize}
    \item $n=6$: See the p6 $\to$ p6 $\to$ p6m example in Fig.~\ref{fig:F1}.
    \item $n=4$: See the p4 $\to$ p4 $\to$ p4m example in Fig.~\ref{fig:F1}.
    \item $n=3$: See the p3 $\to$ p3 $\to$ p3m1 example in Fig.~\ref{fig:SI_patterns}\textbf{b}.
    \item $n=2$: See the pgg $\to$ pgg $\to$ cmm example in Fig.~\ref{fig:F1}.
    \item $n=1$: See the pg $\to$ pg $\to$ pmg example in Fig.~\ref{fig:SI_patterns}\textbf{g}.
\end{itemize}

\textbf{Glide reflectional symmetry}: Gain of glide reflectional symmetry can be observed for all $n = 1,2,3,4,6$:
\begin{itemize}
    \item $n=6$: See the p6 $\to$ p6 $\to$ p6m example in Fig.~\ref{fig:F1}.
    \item $n=4$: See the p4 $\to$ p4 $\to$ p4m example in Fig.~\ref{fig:F1}.
    \item $n=3$: See the p3 $\to$ p3 $\to$ p3m1 example in Fig.~\ref{fig:SI_patterns}\textbf{b}.
    \item $n=2$: See the p2 $\to$ p2 $\to$ p6m example in Fig.~\ref{fig:F4}\textbf{f}.
    \item $n=1$: See the p1 $\to$ p1 $\to$ p31m example in Fig.~\ref{fig:SI_symmetry_gain}\textbf{a}. 
\end{itemize}

\begin{figure}[t!]
    \centering
    \includegraphics[width=\textwidth]{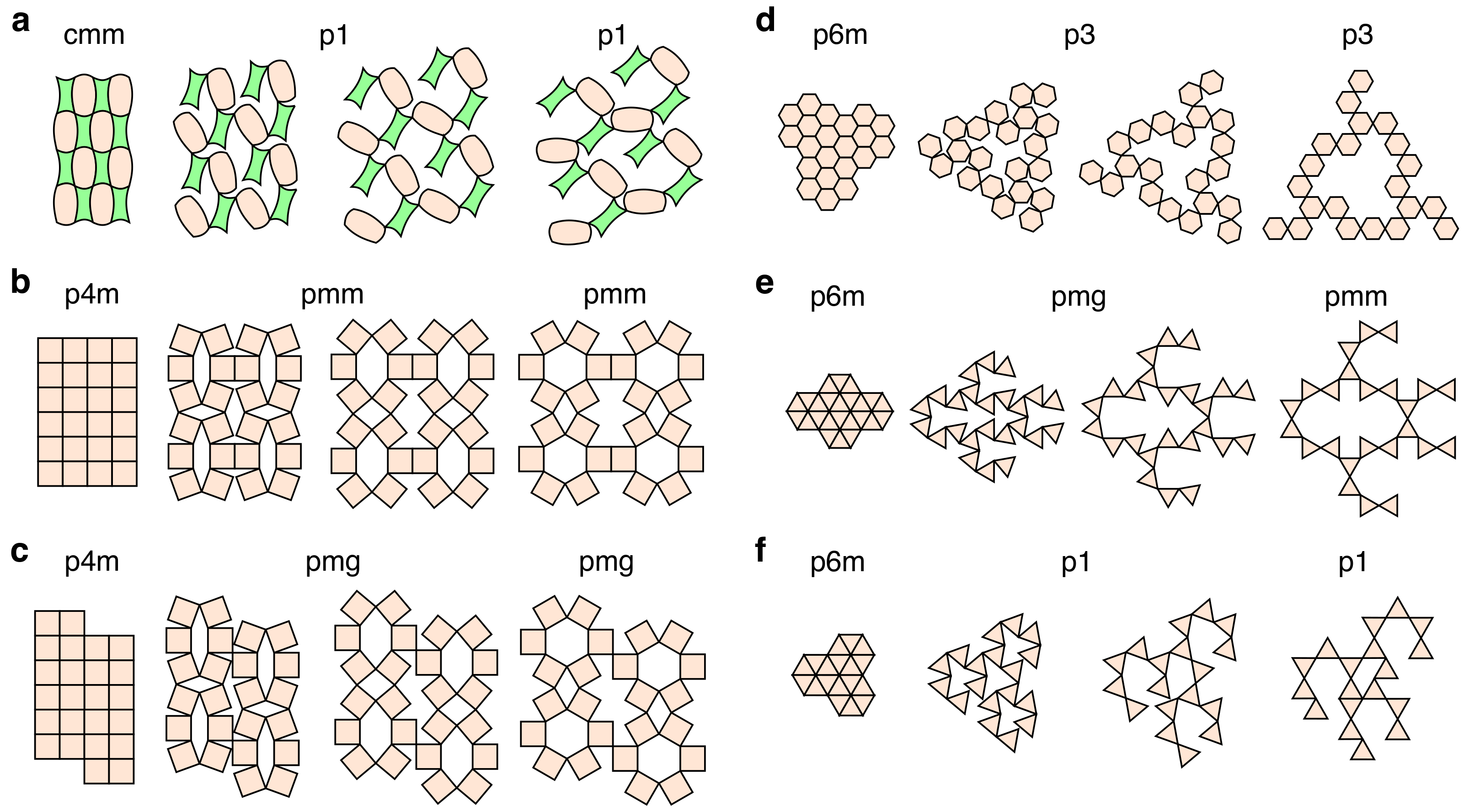}
    \caption{\textbf{More examples of deployable kirigami patterns with rotational symmetry lost throughout deployment.} \textbf{a},~A cmm $\to$ p1 $\to$ p1 example derived from the standard quad pattern. \textbf{b}, A p4m $\to$ pmm $\to$ pmm example derived from the standard quad pattern. \textbf{c}, A p4m $\to$ pmg $\to$ pmg example derived from the standard quad pattern. \textbf{d}, A p6m $\to$ p3 $\to$ p3 example derived from the standard kagome pattern. \textbf{e}, A p6m $\to$ pmg $\to$ pmm example derived from the standard kagome pattern. \textbf{f}, A p6m $\to$ p1 $\to$ p1 example derived from the standard kagome pattern.}
    \label{fig:SI_symmetry_lost}
\end{figure}

\subsubsection{Loss of symmetry}
We can see that an $m$-fold to $n$-fold rotational symmetry loss is possible for any $n,m \in \{1,2,3,4,6\}$ with $n|m$ and $m > n$:
\begin{itemize}
    \item $2\to 1$: See the cmm $\to$ p1 $\to$ p1 example in Fig.~\ref{fig:SI_symmetry_lost}\textbf{a}.
    \item $4\to 2$: See the p4m $\to$ pmm $\to$ pmm example in Fig.~\ref{fig:SI_symmetry_lost}\textbf{b} and the p4m $\to$ pmg $\to$ pmg example in Fig.~\ref{fig:SI_symmetry_lost}\textbf{c}.
    \item $6\to 3$: See the p6m $\to$ p3 $\to$ p3 example in Fig.~\ref{fig:SI_symmetry_lost}\textbf{d}.
    \item $3 \to 1$: See the p31m $\to$ p1 $\to$ p1 example in Fig.~\ref{fig:F4}\textbf{f}.
    \item $6 \to 2$: See the p6m $\to$ pmg $\to$ pmm example in Fig.~\ref{fig:SI_symmetry_lost}\textbf{e}.
    \item $4 \to 1$: See the p4g $\to$ p1 $\to$ p1 example in Fig.~\ref{fig:F5}\textbf{f}.
    \item $6 \to 1$: See the p6m $\to$ p1 $\to$ p1 example in Fig.~\ref{fig:SI_symmetry_lost}\textbf{f}.
\end{itemize}

Loss of reflectional and glide reflectional symmetries can be easily achieved by breaking the connectivity of the tiles (see Section~\ref{sect:lattice} for a more detailed discussion).

\subsubsection{Preservation of symmetry}
It can be observed that preservation of $n$-fold rotational symmetry throughout deployment is possible for all $n = 1,2,3,4,6$:
\begin{itemize}
    \item $n=6$: See the p6 $\to$ p6 $\to$ p6m example in Fig.~\ref{fig:F1}.
    \item $n=4$: See the p4m $\to$ p4g $\to$ p4m example in Fig.~\ref{fig:F1}, and the p4m $\to$ p4 $\to$ p4m example in Fig.~\ref{fig:SI_patterns}\textbf{f}.
    \item $n=3$: See the p31m $\to$ p3 $\to$ p31m example in Fig.~\ref{fig:F1}, and the p3m1 $\to$ p3 $\to$ p3m1 example in Fig.~\ref{fig:SI_patterns}\textbf{d}.
    \item $n=2$: See the cmm $\to$ p2 $\to$ pmm example and the pgg $\to$ pgg $\to$ cmm example in Fig.~\ref{fig:F1}.
    \item $n=1$: See the cm $\to$ p1 $\to$ p1 example and the pm $\to$ cm $\to$ cm example in Fig.~\ref{fig:F1}.
\end{itemize}

Preservation of reflectional and glide reflectional symmetries can also be observed (see the p6m $\to$ p31m $\to$ p6m example and the p4m $\to$ p4g $\to$ p4m example in Fig.~\ref{fig:F1}). 

Furthermore, it is possible to design periodic deployable patterns with $n$-fold rotational symmetry that stay in the same wallpaper group throughout deployment: 
\begin{itemize}
    \item $n=6$: See the p6 $\to$ p6 $\to$ p6 example in Fig.~\ref{fig:SI_patterns}\textbf{a}.
    \item $n=4$: See the p4 $\to$ p4 $\to$ p4 example in Fig.~\ref{fig:SI_patterns}\textbf{o}.
    \item $n=3$: See the p3 $\to$ p3 $\to$ p3 example in Fig.~\ref{fig:SI_patterns}\textbf{e}.
    \item $n=2$: See the p2 $\to$ p2 $\to$ p2 example in Fig.~\ref{fig:SI_patterns}\textbf{i}.
    \item $n=1$: See the p1 $\to$ p1 $\to$ p1 example in Fig.~\ref{fig:F1}.
\end{itemize}

\subsection{Summary of possible symmetry changes}
From the above results on the gain, loss and preservation of rotational symmetry, we have the following theorem:
\begin{theorem} 
For any $n,m \in \{1,2,3,4,6\}$ with $m \geq n$ and $n|m$, it is possible to design a deployable kirigami pattern that achieves an $n$-fold to $m$-fold rotational symmetry change throughout deployment, and a pattern that achieves an $m$-fold to $n$-fold rotational symmetry change throughout deployment. 
\end{theorem}
\noindent \textbf{Proof. } For any such $(m,n)$, we have already constructed an explicit example of deployable kirigami pattern with the desired rotational symmetry change as listed above. \hfill $\blacksquare$\\

Similarly, from the above results on the gain, loss and preservation of reflectional and glide reflectional symmetry, we have the following theorems:
\begin{theorem}
For any $n = 1,2,3,4,6$, it is possible to design a deployable kirigami pattern with $n$-fold rotational symmetry that achieves any target reflectional symmetry change (gain/loss/preservation) throughout deployment.
\end{theorem}
\begin{theorem}
For any $n = 1,2,3,4,6$, it is possible to design a deployable kirigami pattern with $n$-fold rotational symmetry that achieves any target glide reflectional symmetry change (gain/loss/preservation) throughout deployment.
\end{theorem}

\section{Lattice representations}\label{sect:lattice}
\begin{figure}[t]
\centering
\includegraphics[width=\textwidth]{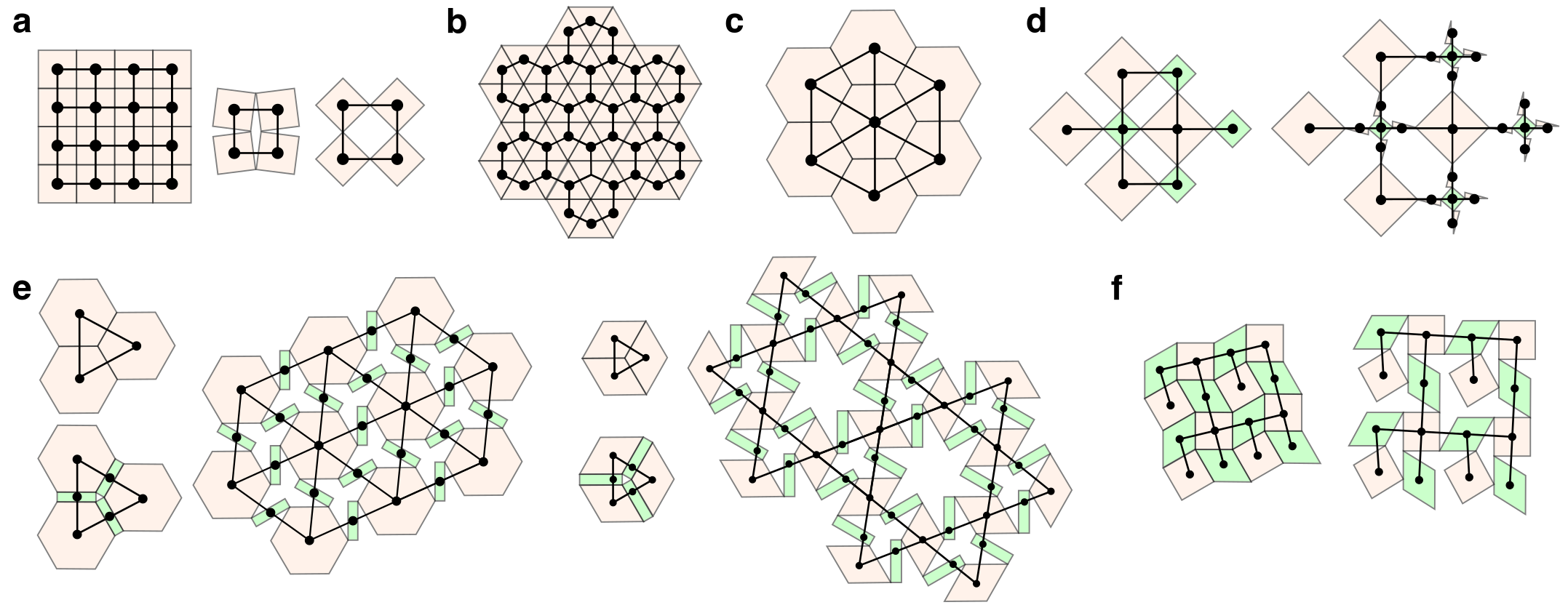}
\caption{\textbf{Lattice representations of kirigami patterns}. \textbf{a}, The regular $(4,4)$ lattice representing the standard quad topology (e.g. the p4m, p4g, pmg patterns in Fig.~\ref{fig:F1}), with each tile connected to exactly four adjacent tiles. \textbf{b}, The regular $(3,3)$ lattice representing the standard kagome topology (e.g. the p6m, p3m1 and p3 patterns in Fig.~\ref{fig:F1}). \textbf{c}, The regular $(6,6)$ lattice. Note that the triangles in it make the structure rigid. \textbf{d}, Using the expansion cuts introduced in Fig.~\ref{fig:F2}, we can turn a deployable structure with the regular $(4,4)$ lattice into another deployable structure with a $(2,4,2)$ lattice. \textbf{e},~Using the expansion tiles introduced in Fig.~\ref{fig:F2}, we can turn the triangles in a rigid lattice into another polygon, thereby producing novel deployable patterns with a $(2,6,2)$ lattice (left) or a $(2,4,2)$ lattice (right). Both examples shown here are p6m $\to$ p6 $\to$ p6. \textbf{f},~Breaking certain connections in a given lattice yields a 1-fold deployable structure with another lattice representation. The example shown here is a p4g $\to$ p1 $\to$ p1 pattern with a $(1,3,4,3,1)$ lattice (see also the p31m pattern in Fig.~\ref{fig:F4}\textbf{f} with a $(2,4,4,2)$ lattice).}
\label{fig:F5}
\end{figure}

\begin{figure}[t]
    \centering
    \includegraphics[width=\textwidth]{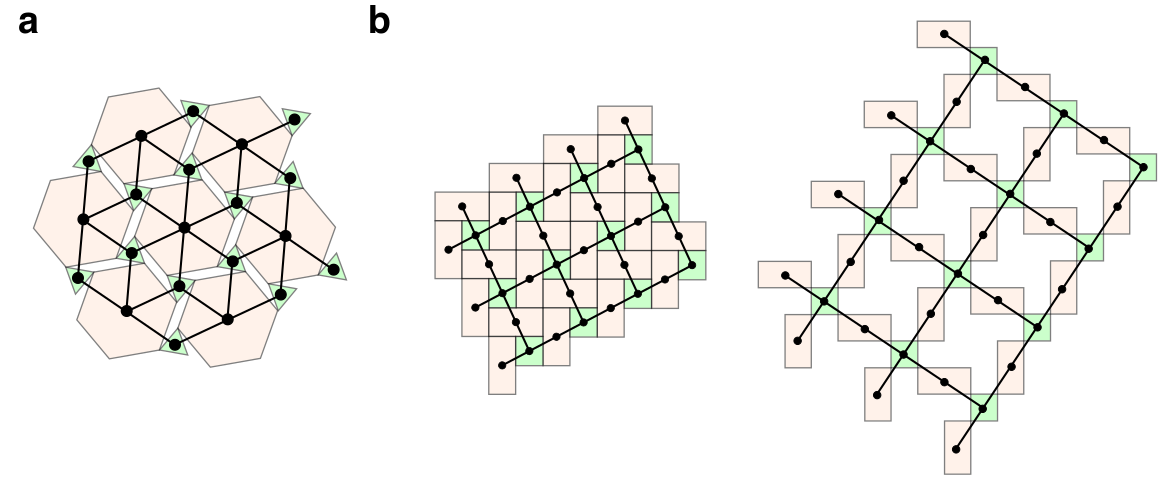}
    \caption{\textbf{More deployable lattice structures.} \textbf{a}, The topology of the p6 pattern in Fig.~\ref{fig:F1} is a deployable $(3,6,3)$ lattice. \textbf{b}, The topology of the p4g and p4 patterns in Fig.~\ref{fig:SI_patterns}\textbf{n}-\textbf{o} is a deployable $(2,4,2)$ lattice.}
    \label{fig:SI_lattice}
\end{figure}

Observing the close relationship between the wallpaper group of a periodic deployable kirigami pattern and its underlying topology, we analyze different patterns in terms of their lattice representations (Fig.~\ref{fig:F5}). In the lattice representation, each tile is represented by a node. An edge between two nodes exists if their corresponding tiles are connected to each other. Here we introduce a cyclic notation for the lattice representation of a periodic kirigami pattern. Starting from a tile with the lowest connectivity, we denote $a_1$ as its number of neighbors. We then consider all neighbors of the tile and choose the one with the lowest connectivity, and denote its number of neighbors as $a_2$. We continue the process until the sequence repeats (i.e. $a_1, a_2, \dots, a_k, a_1, a_2, \dots$), and use $(a_1, a_2, \dots, a_k, a_1)$ to represent the lattice. We remark that while each sequence does not necessarily correspond to a unique lattice structure, it helps us understand the connectivity of any given periodic deployable kirigami pattern. Below, we analyze three lattice types we observed in periodic deployable kirigami patterns (see Fig.~\ref{fig:SI_lattice} for more examples).

\subsection{Regular lattice}
Note that the only regular polygons that can tile the plane are the triangle, square, and hexagon. Therefore, the only regular lattices are $(4,4)$ (Fig.~\ref{fig:F5}\textbf{a}), $(3,3)$ (Fig.~\ref{fig:F5}\textbf{b}) and $(6,6)$ (Fig.~\ref{fig:F5}\textbf{c}). Note that the regular $(4,4)$ lattice and the regular $(3,3)$ lattice correspond to the standard quad kirigami topology and the standard kagome kirigami topology, both of which are deployable. On the contrary, the rigidity of the triangles (3-cycles) in the regular $(6,6)$ lattice prevents it from being deployable. 

\subsection{Augmented lattice}
Another type of lattice we observed can be viewed as an augmented version of the regular lattice, with certain tiles inserted in a rotationally symmetric way. One example is the topology of the p6 pattern in Fig.~\ref{fig:F1}, which is a deployable $(3,6,3)$ lattice. Interestingly, the two symmetry-preserving expansion methods we introduced in Fig.~\ref{fig:F2} provide us with a systematic way of creating new symmetric augmented lattice of deployable structures from any given pattern. For instance, the expansion cuts allow us to turn the regular $(4,4)$ lattice into another deployable structure with a $(2,4,2)$ lattice (Fig.~\ref{fig:F5}\textbf{d}) while preserving the 4-fold rotational symmetry of the lattice. The expansion tiles also effectively add vertices along the edges of the rigid triangles in a given lattice, thereby turning them into other polygons and making the structure deployable, with the rotational symmetry preserved. Fig.~\ref{fig:F5}\textbf{e} shows an example of turning a rigid $(6,6)$ lattice into a deployable $(2,6,2)$ lattice (top), and an example of turning a rigid $(4,4)$ lattice into a deployable $(2,4,2)$ lattice (bottom). This shows that the expansion methods are not only useful geometrically but also mechanically for the design of deployable kirigami patterns. 

\subsection{Trimmed lattice} 
By removing certain connections in the lattice of a given kirigami pattern, we can break the symmetry of the lattice and hence achieve a large variety of changes in the rotational, reflectional or glide reflectional symmetries throughout deployment. For instance, one can obtain a floppy $(1,3,4,3,1)$ lattice as shown in Fig.~\ref{fig:F5}\textbf{f} and a floppy $(2,4,4,2)$ lattice as shown in the p31m pattern in Fig.~\ref{fig:F4}\textbf{f}, which lose all symmetries throughout deployment. This shows that it is possible to create a kirigami pattern in any wallpaper group $G$ with the deployment path $G \to$ p1 $\to$ p1. Using trimmed lattice with carefully designed tile geometries, it is also possible to achieve a symmetry gain such as the 2-fold to 6-fold rotational symmetry gain in the floppy p2 pattern in Fig.~\ref{fig:F4}\textbf{f}.

\section{Graph of possible symmetry changes}
\begin{figure}[t!]
    \centering
    \includegraphics[width=0.9\textwidth]{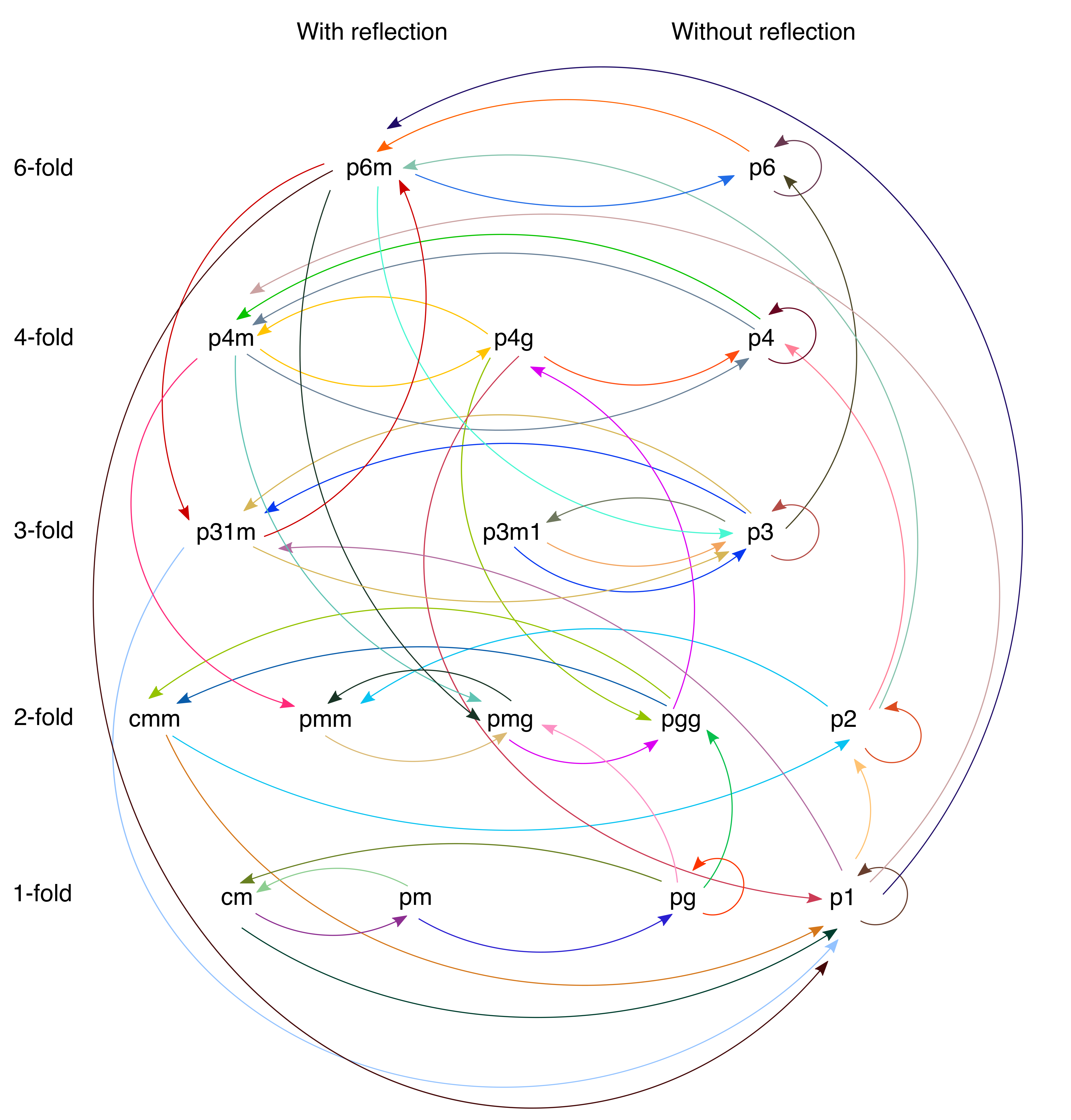}
    \caption{\textbf{The graph $\widetilde{\mathcal{G}}$ of possible group changes observed in the patterns we have identified.} The group change of each example pattern considered in this paper is represented using one distinct color. For instance, the red arrows from p6m to p31m and from p31m to p6m correspond to the p6m $\to$ p31m $\to$ p6m standard kagome pattern in Fig.~\ref{fig:F1}, and the orange arrow from p6 to p6m corresponds to the p6 $\to$ p6 $\to$ p6m pattern in Fig.~\ref{fig:F1}. We remark that $\widetilde{\mathcal{G}}$ is only a subgraph of $\mathcal{G}$, the graph of all possible group changes.}
    \label{fig:SI_graph}
\end{figure}

One can consider the space of the seventeen wallpaper groups as a directed graph $\mathcal{G} = (\mathcal{V}, \mathcal{E})$, where the vertex set $\mathcal{V}$ consists of the seventeen nodes representing the seventeen wallpaper groups, and the directed edge set $\mathcal{E}$ consists of directed arrows indicating all possible group changes throughout deployment of the kirigami patterns. Based on the patterns we have identified in this paper, we construct a subgraph $\widetilde{\mathcal{G}} = (\mathcal{V}, \widetilde{\mathcal{E}})$ of $\mathcal{G}$ where $\widetilde{\mathcal{E}}$ are obtained from the patterns we have identified (see Fig.~\ref{fig:SI_graph}). It is easy to see that $\widetilde{\mathcal{G}}$ is connected, which suggests that $\mathcal{G}$ is also connected.

By Theorem~\ref{thm:existence}, the out-degree of any vertex in $\mathcal{G}$ is at least 1, which is evident from the graph $\widetilde{\mathcal{G}}$. Similarly, by Theorem~\ref{thm:existence_deployed}, the in-degree of any vertex in $\mathcal{G}$ is at least 1. We can easily see that each wallpaper group in the graph $\widetilde{\mathcal{G}}$ is the endpoint of some paths.

By the floppy lattice construction introduced above, the in-degree of p1 in $\mathcal{G}$ should be exactly 17. Note that in the graph $\widetilde{\mathcal{G}}$ in Fig.~\ref{fig:SI_graph}, we have omitted all $G$ $\to$ p1 $\to$ p1 changes except for those explicitly described in the figures in this paper. 

\section{Discussion}
The ability to control the size, perimeter and symmetry changes makes kirigami, long a paradigm for art, an inspiration for the mathematical development of ideas linking (discrete) geometry, topology and analysis, and alluring as the basis for technology in such instances as the design of energy-storing devices~\cite{liu2017flexible}, electromagnetic antennae~\cite{liu2015origami,nauroze2018continuous} etc. Our work has explored the connection between kirigami and the symmetry associated with the planar wallpaper groups. We have shown that it is possible to create deployable patterns using all of the seventeen wallpaper groups, and further studied the size change, symmetry change and lattice structure of these patterns. Many of the results regarding the existence of deployable patterns and the possible symmetry changes in this work are obtained by explicitly constructing different examples. Going beyond the existence of deployable kirigami patterns, the expansion cut method leads to arbitrary size changes, while the expansion tile method creates voids in between some tiles in the contracted state. 

A natural limitation of planar periodic deployable patterns is in their rotational symmetry, with the possible orders of rotation being $n = 1,2,3,4,6$ only. A class of planar tessellations closely related to the wallpaper groups are the aperiodic quasicrystal patterns~\cite{shechtman1984metallic} that can also tile the plane. While lacking translational symmetry, quasicrystal patterns can exhibit rotational symmetry not found in any wallpaper group patterns. Natural next steps include exploring the possibility of using patterns that do not have any voids in the contracted state, creating deployable structures based on quasicrystal patterns, and extending our study of symmetries of deployable patterns to 3D for the design of structural assemblies~\cite{choi2020control}.
\bibliographystyle{ieeetr}
\bibliography{kirigami_wallpaper_groups_bib}

\end{document}